# TeeMAF: A TEE-Based Mutual Attestation Framework for On-Chain and Off-Chain Functions in Blockchain DApps

Xiangyu Liu, Brian Lee, Yuansong Qiao
Software Research Institute, Technological University of the Shannon, Athlone, Ireland
xiangyu.liu.tus@hotmail.com, brian.lee@tus.ie, yuansong.qiao@tus.ie

*Abstract*—The rapid development of Internet of Things (IoT) technology has led to growing concerns about data security and user privacy in the interactions within distributed systems. Decentralized Applications (DApps) in distributed systems consist of on-chain and off-chain functions, where on-chain functions are smart contracts running in the blockchain network, while off-chain functions operate outside the blockchain. Since smart contracts cannot access off-chain information, they cannot verify whether the off-chain functions, i.e. the software components, they interact with have been tampered or not. As a result, establishing mutual trust between the on-chain smart contracts and the off-chain functions remains a significant challenge. To address the challenge, this paper introduces TeeMAF, a generic framework for mutual attestation between on-chain and off-chain functions, leveraging Trusted Execution Environments (TEE), specifically Intel Software Guard Extensions (SGX), SCONE (a TEE container on top of Intel SGX), and remote attestation technologies. This ensures that the deployed off-chain functions of a DApp execute in a provably secure computing environment and achieve mutual attestation with the interacting on-chain functions. Through a security analysis of TeeMAF, the reliability of deployed DApps can be verified, ensuring their correct execution. Furthermore, based on this framework, this paper proposes a decentralized resource orchestration platform (a specific DApp) for deploying applications over untrusted environments. The system is implemented on Ethereum and benchmarked using Hyperledger Caliper. Performance evaluation focusing on throughput and latency demonstrates that, compared to platforms without a mutual attestation scheme, the performance overhead remains within an acceptable range.

*Index Terms*—Blockchain, Trusted Execution Environment (TEE), Intel SGX, Mutual Attestation

## I. INTRODUCTION

The rapid development of Internet of Things (IoT) technology and the advent of the digital era have raised concerns about the security, confidentiality, and integrity of interactions between applications deployed in distributed systems in edge and cloud environments [1]. Data integrity is crucial for the accurate execution of applications, and any data tampering can lead to application failure. Blockchain technology offers essential solutions to these vulnerabilities by ensuring the immutability of transactions and enhancing data transparency [2]. Consequently, numerous proposals suggest incorporating decentralized applications (DApps) in distributed systems [3]. Establishing trust between the blockchain network and the off-chain computational nodes executing DApps remains a critical issue [4]. Smart contracts on the blockchain cannot directly access off-chain data and information sources due to the inherent isolation between the blockchain environment and the external world, which is essential for ensuring the security and data integrity of the blockchain. The current blockchain systems verify the transaction submitter via a signature signed by a private key of the submitter. However, the blockchain systems cannot verify whether the software which is submitting the transaction is trustworthy or not. The software system may be hacked or the user may use a software release with vulnerabilities. As a result, it is extremely challenging for the on-chain functions of a DApp to attest the code integrity of the interacting off-chain functions.

Oracles are frequently used in blockchain networks to access off-chain data. The reliability of these oracles is a significant issue. If an oracle is unreliable or compromised, it can introduce untrustworthy external data into the blockchain [5]. Solutions like Provable [6] and ChainLink [7] improve data credibility by using multi-signatures and distributed nodes to address this problem. However, these solutions only verify the authenticity and intactness of the data entering the oracle, failing to attest the authenticity and intactness of the off-chain information sources transmitting the data, which results in a lack of complete end-to-end trustworthiness.

In off-chain trusted computations, there is growing interests in utilizing secure Multi-Party Computation (MPC) [8], Zero-Knowledge Proofs (ZKP) [9] technologies. MPC is a powerful approach that enables multiple parties to collaboratively compute on their respective inputs while preserving input privacy. This methodology enables secure data analysis collaborations without sacrificing the confidentiality of the data involved [10]. ZKP technology is especially useful in blockchain settings, as it allows for the validation of transactions or data states without revealing the underlying data or transaction details, thus improving privacy [11], [12]. ZKP plays a crucial role in maintaining the confidentiality and integrity of transactional data. Nevertheless, both approaches have inherent limitations. MPC faces performance and scalability issues, while ZKP is associated with high complexity and significant resource consumption. These limitations prevent them from independently addressing all trust-related challenges [13].

Trusted Execution Environments (TEE) [14] technology enables the creation of isolated execution environments within processors, protecting sensitive code and data during processing [15]. TEEs are essential for off-chain computations requiring enhanced security [16]. It ensures the protection of data integrity and confidentiality, particularly during the execution of sensitive operations, thus offering a comprehensive solution to the challenges of maintaining privacy and security in off-chain computing [17].

Therefore, TEE technologies offer an alternative approach to address the trust issues in the interaction between the off-chain functions of DApps running on off-chain computation nodes and the on-chain functions of DApps running on the blockchain. This study specifically adopts Intel SGX [18], a TEE technology with a core mechanism known as remote attestation. Remote attestation enables third parties to verify the identity, intactness, and secure execution of applications running within an enclave on Intel SGX-supported platforms.

In summary, to address the trust issues between the on-chain and off-chain components of DApps deployed in distributed systems and prevent malicious tampering of the code implementing off-chain functionalities, this paper proposes a mutual attestation framework incorporating blockchain and TEE such that the smart contracts and the offchain functions of a DApp can attest each other to ensure that they are interacting with the original (intact) functions. This framework is referred to as the TEE-Based Mutual Attestation Framework (TeeMAF). The main contributions of this paper are as follows:

1) A novel framework is proposed that combines blockchain and TEE to securely bind and mutually attest the functionalities of on-chain and off-chain components within a DApp, thereby enhancing overall system trust and security.
2) Based on the TeeMAF framework, a decentralized resource orchestration platform (named DROP) is proposed. A set of experiments are conducted to validate the feasibility of TeeMAF and the orchestration platform, and to evaluate the performance including transaction throughput and latency. The experimental results demonstrate that TeeMAF maintains robust performance while ensuring security, trustworthiness, and confidentiality.

The remainder of this paper is organized as follows: Section II introduces the background knowledge of blockchain technology and TEE. Section III describes the proposed framework. Section IV presents a security analysis of the framework. Experimental results and performance evaluations are discussed in Section V. Section VI discusses related works. Finally, Section VII concludes the work presented in this paper.

II. BACKGROUND

This section will briefly introduce the key concepts that the TeeMAF framework relies upon: blockchain and TEE.

*A. Blockchain and Smart Contract*

Blockchain, a distributed ledger technology, is employed by prominent cryptocurrencies like Bitcoin and Ethereum. By allowing all participants to collaboratively maintain the ledger's integrity and consistency, this technology eliminates the need for a trusted intermediary, thereby achieving decentralization and mitigating potential points of failure and fraud [19]. Blockchain technology organizes information into blocks, with each block containing a set of transactions connected to the preceding block. These transaction records are open and transparent to all participants, allowing anyone to view and verify the transactions. Each block includes the hash value of the preceding block, forming a continuous chain. This structure guarantees immutability and transparency [20]. Once data is recorded on the blockchain, it is documented by all nodes, making it difficult to alter.

Nick Szabo [21] first introduced the concept of smart contracts in 1997, defining them as computer programs embedded within computerized transaction protocols to automate the execution of specific contractual tasks. These contracts are designed primarily to ensure enforcement while preventing malicious actions and unforeseen contingencies. In recent years, smart contracts have rapidly evolved with the development of blockchain platforms like Ethereum. The decentralized and immutable nature of smart contracts enables automated execution and verification without third-party intervention. Moreover, the widespread adoption of smart contracts has catalyzed the growth of decentralized applications (DApps). Unlike traditional applications, DApps operate on a decentralized network without a single point of control, enhancing security and transparency [22]. They are driven by the core logic of smart contracts and involve both on-chain and off-chain functions. On-chain functions refer to operations executed within the smart contracts on blockchain, maintained and verified by the network's nodes, ensuring transparency and immutability. However, on-chain operations often face limitations in terms of efficiency and cost. Off-chain functions, which occur outside the blockchain, can include data processing, computational tasks, and interactions with external systems. By relocating certain computations and data storage off-chain, the system's efficiency and scalability can be significantly improved [23].

*B. Trusted Execution Environments*

A Trusted Execution Environment (TEE) is an isolated domain typically situated within a secure area of the main processor. It operates concurrently with the operating system, ensuring the confidentiality and integrity of code and data even in scenarios where the operating system is compromised or deemed untrustworthy.

*Intel Software Guard Extensions:* Introduced by Intel in 2013, Intel SGX is an instruction set extension designed to provide a Trusted Execution Environment in user space, known as an Enclave, with mandatory hardware security independent of the security status of firmware and software. Each SGX-enabled processor has two critical cryptographic keys embedded in its hardware fuses: the Root Provisioning Key (RPK) and the Root Sealing Key (RSK). These keys are inaccessible and tamper-resistant from external access, thereby enhancing security. The RPK is randomly generated and retained by a dedicated offline key generation device at Intel, forming the basis for a processor to authenticate itself as a genuine Intel SGX CPU within a specific Trusted Computing Base (TCB). The RSK is generated during the manufacturing process, and Intel does not retain a copy of this key. The Enclave employs derived keys from the RSK to encrypt data and ensure its integrity [24].

*Remote Attestation:* Intel SGX provides a mechanism for remote attestation of enclave identities. Third parties can verify an enclave's identity and securely provide it with keys, credentials, and other sensitive data. This remote attestation of an enclave is facilitated by a dedicated security component known as the Quoting Enclave (QE) residing in the same platform with the enclave under attestation [25]. The QE employs a platform-specific asymmetric Attestation Key (AK), which is derived from the RPK, to receive security reports generated by other enclaves. The QE then converts these reports into a signed format called a Quote. During the operation of the enclave, only the QE has access to the AK, and the AK is bound to the version of the processor firmware. Consequently, the Quote can be considered as issued by the processor itself. This Quote can be utilized by remote attestation services to confirm the identity and integrity of the sender enclave [26].

*SCONE:* SCONE [27] is a platform that leverages Intel SGX technology to build and run secure applications. It functions as a secure container mechanism within Docker, utilizing Intel SGX to safeguard container processes from external attacks. SCONE facilitates the transformation of native application images into SCONE confidential application images, enabling the encryption of every service within the Enclave. This capability ensures that applications operate securely by protecting sensitive data and operations within the confines of trusted execution environments provided by Intel SGX.

## III. PROPOSED FRAMEWORK

This paper proposes TeeMAF, a framework based on Trusted Execution Environment (TEE) technologies, designed to facilitate mutual attestation between on-chain and off-chain functionalities within deployed blockchain DApps. The details of the proposed framework are as follows:

### A. Design Goals

*Facilitate Trust Between On-Chain and Off-Chain Function Interactions:* In blockchain systems, a user signs a transaction with their private key to prove the ownership and authorize the transaction. If the transaction contains a smart contract call, the correspondent functions will be executed. However, the blockchain nodes (miners) can only verify that this transaction is signed by a specific private key, they cannot verify whether the signature is performed by trusted software, e.g. the software may be hacked, or the user may send the transactions from different software, and some may contain bugs. TeeMAF aims to provide mutual trust between the smart contracts and the off-chain functions of a DApp. While implementing trust from off-chain functions to smart contracts is trivial (e.g. by hardcoding the smart contract addresses in the off-chain software), this paper focuses on the mechanism to enable attestation from smart contracts to off-chain functions.

To validate TeeMAF, this paper proposes a specific DApp (i.e. containing smart contracts and offchain functions) for computational resource orchestration, which acts as a broker so that resource owners can register their resources in the platform, and DApp owners can rent resources from the platform and deploy their functions on these resources. Both the resource management platform and those DApps deployed on the platform use TeeMAF to achieve mutual attestation between on-chain and off-chain functions. The goals of this platform include:

1) *Offload Task To Trusted Off-Chain Workers in Untrusted Environments:* To enhance security and manage computational loads efficiently, the platform (a DApp) enables generic DApps to deploy computational intensive tasks and data storage to trusted workers running in third-party untrusted off-chain computing nodes.
2) *Controlled Disclosure of Private Information:* Enable the DApp owners (both the resource orchestration platform, and the platform users) to selectively disclose specific information in the on-chain and off-chain interaction processes, while keeping other details confidential, thus achieving selective privacy.

### B. Architecture

In the TeeMAF framework, one of the primary challenges of the attestation protocol is to achieve mutual attestation between on-chain and off-chain functionalities. This challenge is particularly evident when attempting to enable on-chain functions to attest to their linked off-chain functions. This challenge arises primarily due to the inherent limitations of on-chain code, which cannot directly access external systems or independently verify the authenticity of off-chain data without relying on trusted oracles or off-chain computation solutions.

The key idea of TeeMAF is to securely bind a private key with the off-chain functions using the TEE mechanisms so that only the function running inside a TEE enclave can access the private key. In this way, the smart contract can verify the signature signed by the private key to ensure the message is from the correspondent function (running in a TEE enclave).

The architecture of the TeeMAF framework is illustrated in Fig. 1. The framework consists of seven entities: DApp Owner, Blockchain, SCONE Container, Attestation Service, SCONE Local Attestation Service (LAS) and SCONE Configuration and Attestation Service (CAS), The role of each entity in the architecture is clearly defined below:

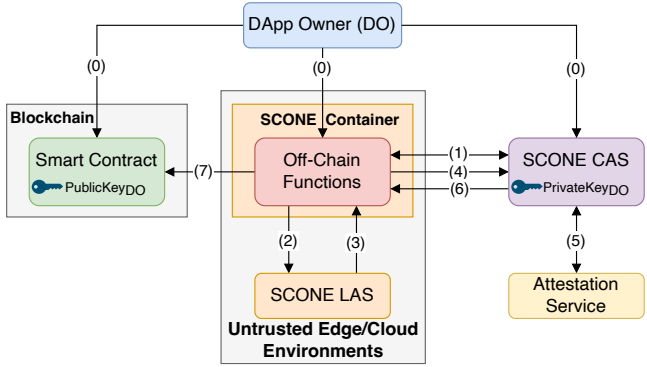

Fig. 1. TeeMAF Architecture

*DApp Owner*: The DApp Owner (DO) is the administrator and creator of a DApp, which is divided into on-chain smart contracts and off-chain functionalities that execute outside the blockchain.

*Blockchain:* Blockchain is a distributed ledger for recording information. It includes the DO's smart contracts, which encapsulate the on-chain functions of the $DApp_{DO}$.

*SCONE Container:* A SCONE container is instantiated from a SCONE confidential application image. It includes the off-chain functions of the $DApp_{DO}$.

*Attestation Service:* The Attestation Service within Intel SGX includes two main types: Intel Attestation Services [24] and Intel Data Centre Attestation Primitives (DCAP) [25]. IAS uses the Intel Enhanced Privacy ID (EPID) methodology for broad applicability in diverse environments, facilitating software attestation within SGX enclaves. Conversely, DCAP utilizes the Elliptic Curve Digital Signature Algorithm (ECDSA), providing more flexibility and scalability for data center environments and enabling organizations to manage attestations according to their specific security needs.

*SCONE Local Attestation Service (LAS)*: The SCONE LAS introduces an independent Quoting Enclave for generating QUOTES.

*SCONE Configuration and Attestation Service (CAS):* A service component of the SCONE infrastructure operates within enclaves in a cloud environment and is responsible for managing security policy documents containing confidential information about DO. It is responsible for collecting attestation collaterals and coordinating with the Attestation Service in order to attest entities requesting access to confidential information. Only services that have been given explicit permission by the application policy can access confidential information after successful attestation and verification.

*Workflow*: The high-level workflow of TeeMAF is shown in Fig. 1. The steps in Fig.1 are described below. A more detailed sequence diagram is provided in the next subsection.

0) This step mainly involves some initialization work by the DO, including deploying the smart contract of $DApp_{DO}$ to the blockchain, generating a public/privat key pair representing the offchain functions, registering the public key (representing the off-chain functions) to the smart contract, converting the native image of the off-chain functions into a SCONE confidential application image, instantiating the container, deploying $Policy_{DO}$ in SCONE CAS containing confidential information that will be injected to the off-chain functions (SCONE Container) later, e.g. the private key (representing the off-chain function) and the smart contract address.

1) Before the off-chain functions interact with the on-chain smart contract, they first need to request the confidential information in $Policy_{DO}$ from the SCONE CAS, and then the SCONE CAS needs to request attestation collaterals (QUOTE).

2) The off-chain functions call the *EREPORT* instruction to generate a report and send it to the SCONE LAS on the same platform to request a signature to generate the QUOTE.

3) The SCONE LAS performs local attestation on the Enclave where the off-chain functions are located to ensure they are on the same platform, then signs the report using the Attestation Key to obtain the QUOTE and returns it.

4) The off-chain functions send the QUOTE to the SCONE CAS.

5) The SCONE CAS sends the QUOTE to the Attestation Service and generates an Attestation Verification Report in return.

6) If the Attestation Verification Report is positive, the Enclave where the off-chain functions are located has completed a remote attestation. The SCONE CAS will inject the private key and the smart contract address into the off-chain functions in the SCONE container, according to $Policy_{DO}$.

7) Every time the off-chain functions interact with the on-chain smart contract, they will carry a signature with a hash value generated by a random number and signed with the private key, so that the on-chain smart contract can verify that it is interacting with owner defined off-chain functions.

C. *TeeMAF Detailed Design*

This section provides the detailed design of the framework and related protocols.

TABLE I
ABBREVIATION DESCRIPTION

| Abbreviation | Description |
|---|---|
| $OFFCF$ | The DApp's Off-Chain Functions |
| $ONCF$ | The DApp's On-Chain Functions |
| $PK$ | The Secp256k1 PublicKey |
| $SK$ | The Secp256k1 PrivateKey |
| $EA$ | The Ethereum Address |
| $SID$ | The Service ID |
| $MRE$ | The MRENCLAVE |
| $SC$ | Smart Contract |
| $SCA$ | Smart Contract Address |
| $Sig$ | Signature |
| $Msg$ | Message |

Table I lists the abbreviations for the entities and parameters essential to the mutual attestation protocol in the TeeMAF Framework. Fig 2. illustrates the sequence diagram of the mutual attestation protocol for the interaction of on-chain and off-chain functions in a DApp. The attestation protocol consists of three distinct phases: Initialization, Attestation and Injection, and Execution and Verification.

1) *Initialization:* Generally, a DApp consists of two parts: on-chain functions and off-chain functions. On-chain functions specifically refer to the business logic functions in the DApp Smart Contract ($SC_{DApp}$) that can alter the blockchain state and interact with off-chain functions, as defined by the DApp Owner (DO). Initially, the DO generates a public key ($PK_{OFFCF}$) and private key ($SK_{OFFCF}$) for the DApp's off-chain functions using the Secp256k1 algorithm. The $PK_{OFFCF}$ is then converted into an Ethereum Address ($EA_{OFFCF}$) using the Keccak-256 cryptographic function. Subsequently, the DO calls the RegisterPK() method in the $SC_{DApp}$, passing $EA_{OFFCF}$ as an argument to register the public key of $DApp_{OFFCF}$ on the blockchain. This method can only be invoked by the DO, who deployed $SC_{DApp}$, to register the public key. Furthermore, the DO uses the SCONE toolchain to transform the $DApp_{OFFCF}$ image into a SCONE confidential image, determining the Service ID ($SID_{OFFCF}$) and the MRENCLAVE ($MRE_{OFFCF}$) of the enclave for this image. Finally, the DO creates a Policy according to the SCONE security policy specifications, incorporating relevant information about $SCA_{DApp}$, $SK_{OFFCF}$, $SID_{OFFCF}$, and $MRE_{OFFCF}$, and registers this Policy with the SCONE CAS.

2) *Attestation and Injection:* When the SCONE confidential image of a DApp's off-chain function is instantiated as a container ($Container_{OFFCF}$), the off-chain function requests the necessary confidential information from the SCONE CAS to interact with the on-chain function. However, before providing the confidential information, the SCONE CAS must perform remote attestation on the $Container_{OFFCF}$ to verify its identity, intactness, and secure execution within an enclave on an Intel SGX enabled platform. First, the SCONE CAS identifies the $SID_{OFFCF}$ and $MRE_{OFFCF}$ to validate the identity and integrity of the $Container_{OFFCF}$ requesting the confidential information, ensuring that the requester is bound by the corresponding security policy. Since the SID in the numerous security Policies registered in the SCONE CAS are unique, the $Container_{OFFCF}$ can match the appropriate security policy when communicating with the SCONE CAS. Subsequently, the MRE computed during the $Container_{OFFCF}$'s launch is compared to determine the program's integrity. The SCONE runtime then calls the *EREPORT* instruction to send the REPORT of the enclave hosting the off-chain function ($Enclave_{OFFCF}$) to the SCONE LAS for signing. The Quoting Enclave [25] in the SCONE LAS first attests that the $Enclave_{OFFCF}$ is on the same platform as itself. After local attestation, it signs the REPORT using a private key derived from the Intel SGX CPU's Root Provisioning Key(RPK), generating a QUOTE that can be verified by a attestation service. Subsequently, SCONE LAS forwards these QUOTES to $Container_{OFFCF}$, which in turn forwards them to SCONE CAS. SCONE CAS then forwards the QUOTES to the Attestation Service for verification, obtaining a attestation verification report. If the report is positive, the SCONE CAS completes the remote attestation of the $Enclave_{OFFCF}$ and injects the necessary confidential information into the $Container_{OFFCF}$.

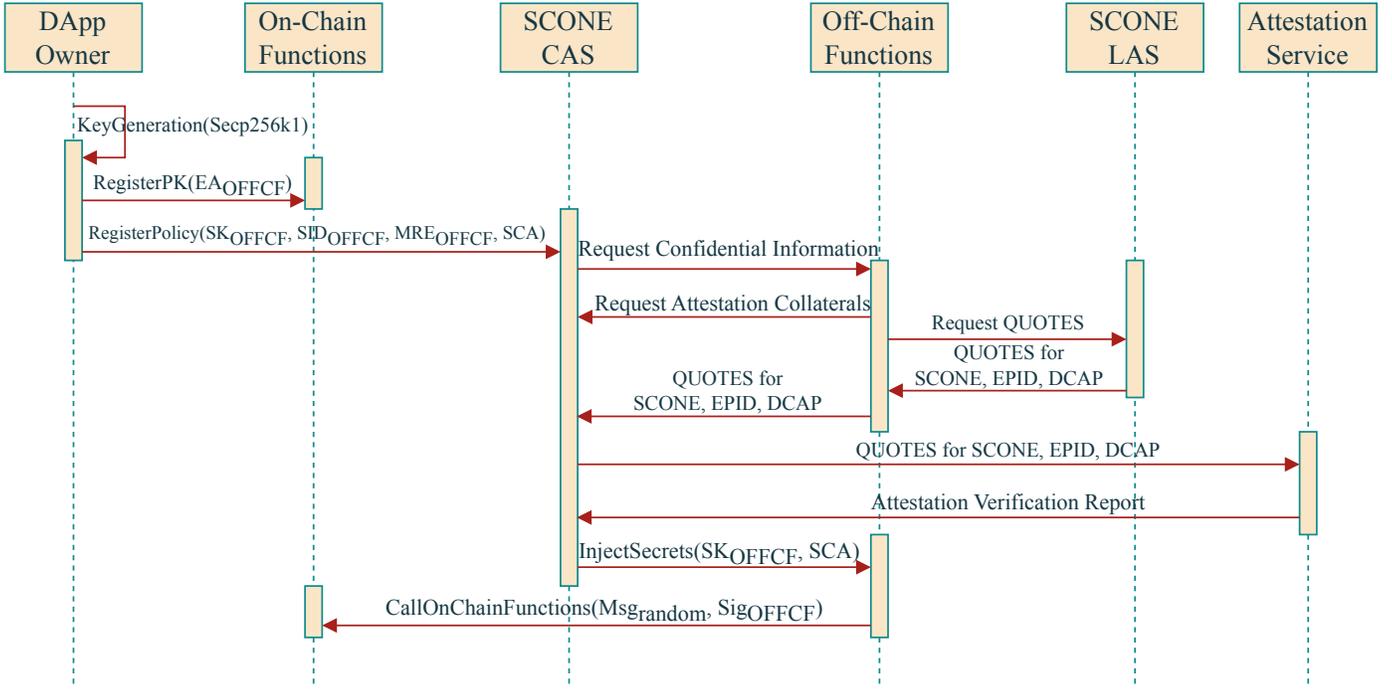

Fig. 2. TeeMAF Mutual Attestation Protocol Sequence Diagram

3) *Execution and Verification:* Once the $Enclave_{OFFCF}$ in the $Container_{OFFCF}$ successfully completes remote attestation through the SCONE CAS, the SCONE CAS injects the confidential information from the Policy into the container, including the critical $SK_{OFFCF}$ and $SCA_{DApp}$. $SK_{OFFCF}$ is the Secp256k1 algorithm private key generated by the DO for the off-chain function during the Initialization Phase, while $SCA_{DApp}$ is the smart contract address where the on-chain function of the DApp interacting with the off-chain function resides. At this point, the off-chain function's attestation of the on-chain function is complete because $SCA_{DApp}$ is injected into the $Container_{OFFCF}$ by the SCONE CAS after the off-chain function has completed remote attestation. Every time the off-chain function interacts with the on-chain function, the OffChainFunctionSignatureVerify() method in the smart contract is invoked. First, the $Container_{OFFCF}$ generates a $Msg_{Random}$ during each interaction and signs it using $SK_{OFFCF}$ to obtain $Sig_{OFFCF}$. Then, $Msg_{Random}$ and $Sig_{OFFCF}$ are passed as parameters into the OffChainFunctionSignatureVerify() method. The hash value of $Msg_{Random}$ ($Hash_{Msg}$) is calculated, and an Ethereum address is recovered using the ECDSA.Recover() method with $Hash_{Msg}$ and $Sig_{OFFCF}$ as parameters. The recovered Ethereum address is compared with the registered public key $EA_{OFFCF}$ in $SCA_{DApp}$. If the two values are verified to be the same, it indicates that the DApp's on-chain function has successfully attested the off-chain function. The attestation algorithm is shown in Algorithm 1.

**Algorithm 1:** Attestation Protocol

**Require:** $PK_{OFFCF}, SK_{OFFCF}, SID_{OFFCF}, MRE_{OFFCF}, SCA_{OFFCF}$
**Ensure:** isAttested ∈ {0, 1}

1: $EA_{OFFCF} \leftarrow$ Keccak256($PK_{OFFCF}$)[-20:]
2: $SCA_{DApp} \leftarrow$ RegisterDAppPublicKey($EA_{OFFCF}$)
3: $Policy_{OFFCF} \leftarrow$ RegisterPolicy($SK_{OFFCF}, SID_{OFFCF}, MRE_{OFFCF}, SCA_{DApp}$)
4: isCASVerified $\leftarrow$ CAS.Verify($SID_{OFFCF}, MRE_{OFFCF}$)
5: **if** isCASVerified == true **then**
6:   QUOTE $\leftarrow$ GatherAttestationCollateral()
7:   isVerified $\leftarrow$ IAS || DCAP.Verify(QUOTE)
8:   **if** isVerified == true **then**
9:     $SK_{OFFCF}, SCA_{DApp} \leftarrow$ CAS.InjectSecrets($Policy_{OFFCF}$)
10:     $Sig_{OFFCF} \leftarrow$ Sign($Msg_{random}, SK_{OFFCF}$)
11:     $Hash_{Msg} \leftarrow$ ECDSA.toEthSignedMessageHash($Msg_{random}$)
12:     recoveredAddress $\leftarrow$ ECDSA.recover($Hash_{Msg}, Sig_{OFFCF}$)
13:     **if** recoveredAddress == $EA_{OFFCF}$ **then**
14:       isAttested $\leftarrow$ 1
15:     **else**
16:       isAttested $\leftarrow$ 0
17:     **end if**
18:   **else**
19:     isAttested $\leftarrow$ 0
20:   **end if**
21: **else**
22:   isAttested $\leftarrow$ 0
23: **end if**

D. *Decentralized Resource Orchestration Platform*

The architecture of the Decentralized Resource Orchestration Platform (DROP) developed based on the TEE-based Mutual Attestation Framework (TeeMAF) is illustrated in Fig. 3. This platform serves as a practical use case for implementing mutual attestation in the interaction between on-chain and off-chain functions of a blockchain DApp. As a DApp itself, the platform utilizes TeeMAF to establish trust between the off-chain compute resource manager (Service Provider's off-chain functions) and the smart contract (Service Provider's on-chain functions). These two components collaboratively manage the compute resources provided by untrusted third parties. Users of the DROP platform can deploy their own DApps on the untrusted compute resources and achieve mutual authentication during the interaction between on-chain and off-chain functions. Essentially, DROP users need to deploy their own smart contracts on the blockchain, rent resources (workers) from the DROP platform, deploy their off-chain functions on the workers, and then use TeeMAF to realize mutual authentication between the on-chain and off-chain functions in the user's DApp. The platform consists of five components: Blockchain, Decentralized Storage, Off-Chain Worker, SCONE CAS and Attestation Service. The role of each entity in the architecture is defined below:

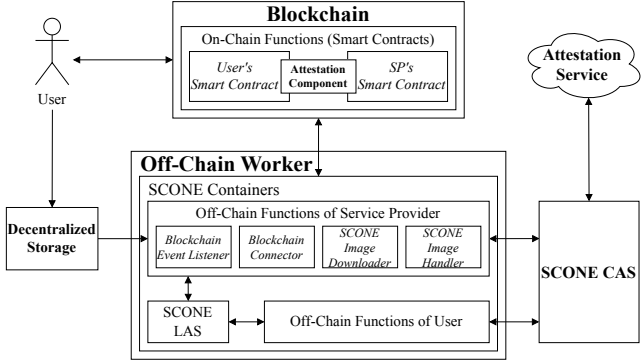

Fig. 3. DROP Architecture

*Blockchain*: In DROP, the core component within the blockchain module is the Service Provider's smart contract ($SC_{SP}$) in the smart contract module. Its primary functions include: OnChainAttestation() in the attestation component, This function, composed of a generic attestation algorithm, implements the remote attestation of off-chain functions by on-chain functions in TeeMAF. It manages the Secp256k1 public keys of all off-chain functions and receives the corresponding public keys based on the requirements of different off-chain functions. Any on-chain operation that requires modifying the blockchain state must invoke this function to ensure successful execution. OffChainWorkerRegister(), this function is used to register off-chain computational nodes that support Intel SGX as off-chain workers on the blockchain, making them available execution entities. OffChainWorkerOrchestration(), this function is responsible for receiving the SCONE confidential application image labels required by $DApp_{User}$ and deploying them to the user-specified off-chain worker.

*Decentralized Storage*: Decentralized Storage (DS) component operates through the blockchain network, offering a robust alternative to traditional centralized data storage solutions. By distributing data across numerous nodes instead of centralizing it under a single entity, DS enhances data security and redundancy, effectively reducing single points of failure and resisting censorship or tampering. The purpose of DS is to manage the SCONE confidential application images of off-chain functions for users and SP. This paper uses InterPlanetary File System (IPFS), a distributed peer-to-peer file storage and sharing system.

*Off-Chain Worker*: Off-chain workers serve as the execution environment for the instantiation of SCONE confidential application images of the off-chain functions of users' and SP's DApps. The off-chain worker component consists of the off-chain functionalities of $DApp_{SP}$, which connect to the blockchain and listen for worker orchestration events. It receives the SCONE confidential image labels submitted by users, retrieves the images from DS, and instantiates them for execution. The component continuously monitors the state of the instantiated containers, and when a container stops or exits, it can interact with $SC_{SP}$ to report the container state information.

*SCONE CAS*: SCONE CAS runs within an enclave and is responsible for managing secure policy documents that contain the confidential information of users and SP. Users and SP have the option to deploy their own SCONE CAS or utilize the SCONE CAS provided by cloud service providers.

*Attestation Service*: Same as described in the TeeMAF framework.

*Workflow*: Both the SP and Users need to perform initialization operations in the TeeMAF framework. The off-chain functions of $DApp_{SP}$ attest the capability of the off-chain worker's TEE (specifically Intel SGX) by completing remote attestation, then register it on the blockchain and listen for on-chain events. The on-chain functions ($SC_{SP}$) of $DApp_{SP}$ manage the information of off-chain workers and orchestrate resources for tasks submitted by users. For the User, after converting the native image of $DApp_{User}$'s off-chain functions into a SCONE confidential image, it needs to be uploaded to any decentralized storage. The user first calls the OffChainWorkerLookUp() method in $SC_{SP}$ to find available off-chain workers. Then, the user calls the OffChainFunctionDeploy() method to submit an application, with the CID of $SCONEImage_{User}$ as input. Upon receiving the deployment request, the off-chain worker retrieves the SCONE confidential image of the user's off-chain functions from the decentralized storage and creates a SCONE container instance. After the user's off-chain functions complete all interactions with the on-chain functions ($SC_{User}$), the off-chain worker running the instance reports the task completion status to $SC_{SP}$.

IV. SECURITY ANALYSIS

In designing the TeeMAF, the contributions primarily focus on achieving mutual attestation between the on-chain and off-chain functionalities within a DApp, ensuring that the code of the off-chain functions remains unaltered. The framework relies on blockchain technology, and TEE, particularly Intel SGX. This section presents a security analysis of TeeMAF, describing the trust assumptions and threat model.

*A. Trust Assumptions*

This paper makes several assumptions regarding blockchain and Intel SGX:

*Immutability and Tamper-Resistance*: It is assumed that the blockchain ledger is immutable and tamper-resistant. Once a transaction or data is recorded on the blockchain, it is considered permanent and unalterable. Any attempt to modify or delete previously recorded information will be detected and rejected by the consensus mechanism.

*Smart Contracts*: The thesis assumes that the smart contracts deployed on the blockchain are secure, reliable, and execute as intended. It is assumed that the programming language used for writing smart contracts, such as Solidity, is well-designed and provides the necessary features for implementing complex logic and business rules. The execution of smart contracts is assumed to be deterministic and consistent across all nodes in the network, ensuring the integrity and transparency of the automated transactions and agreements.

*Hardware-based Isolation*: Intel SGX provides a secure and isolated execution environment known as an Enclave. The Enclave is protected by hardware-based mechanisms that ensure the confidentiality and integrity of the code and data within it. The thesis assumes that the hardware-based isolation provided by SGX is robust and resistant to tampering, even in the presence of privileged software or physical attacks.

*Remote Attestation Mechanism*: The thesis assumes that the attestation mechanism provided by Intel SGX is secure and reliable. Attestation allows remote parties to verify the integrity and authenticity of the code running inside an Enclave. It is assumed that the attestation process is tamper-proof and provides a strong assurance that the Enclave code has not been modified or compromised.

*Side-Channel Resistance*: The thesis assumes that Intel SGX is resistant to certain side-channel attacks (Xu et al., 2015), such as cache timing attacks or power analysis attacks. It is assumed that the hardware-based protections and the careful design of the SGX architecture mitigate the risk of information leakage through side channels, ensuring the confidentiality of the Enclave's sensitive data.

### B. Threat Model

The threat model underlying the TeeMAF presented in this thesis is based on Intel's adversary model [28]. In addition to the standard adversary model, TeeMAF extends the threat model by incorporating a rollback adversary [29], similar to the one described in the SCONE framework.

*System Software Adversary*: This paper assumes that an attacker may possess administrator credentials, granting them access to the root directory of the system software. Furthermore, the attacker may coerce the administrator to duplicate or modify data or code at the system or application level. One of the significant advantages of Intel SGX technology is its ability to defend against not only external attackers but also internal attackers with root access privileges. Since the Enclave memory is encrypted and protected from external access, the attacker cannot find any sensitive information in the stack. However, as the binary files are not encrypted, a copy of the original binary that launches the Enclave remains in the main memory outside the Enclave. An attacker may attempt to obtain sensitive information by analyzing these binary files. To mitigate this risk, confidential data can be securely passed into the Enclave after remote attestation through the SCONE CAS (Configuration and Attestation Service). This ensures that the attacker cannot retrieve the sensitive data within the Intel SGX Enclave.

*Network Adversary*: This paper assumes that an adversary with network access may attempt to communicate with confidential services to gain access or trigger errors within those services. The adversary has the capability to connect to and control various network architectures, including the platform, intranet, internet, and other platform resources. Furthermore, the attacker can communicate with and manipulate remote systems through specified APIs. To mitigate these risks, TeeMAF employs the SCONE network shield to restrict the network communication of applications. By enabling the network shield functionality in the policies of the SCONE Configuration and Attestation Service (CAS), only explicitly permitted network communication is allowed. This mechanism provides TLS-based mutual authentication and encryption, ensuring confidentiality, integrity, and controlled access.

*Simple Hardware Adversary*: This paper assumes that the adversary possesses the capability to remove Dynamic Random-Access Memory (DRAM), has access to hardware debugging tools, and can monitor bus transfers. Additionally, the attacker may obtain decommissioned hardware. Upon hardware or service decommissioning, the attacker can recover data from the CPU, Trusted Platform Module (TPM), and databases. To address these potential threats, TeeMAF relies on the Intel SGX Enclave technology, which effectively defends against such attacks.

*Roll-Back State Adversary*: This paper assumes that the adversary has the capability to roll back the state of the disk, which may contain encrypted databases with updated keys. Furthermore, the attacker may attempt to extract information from the encrypted key database by rolling back the CPU firmware or the confidential service's code, installing an older version of the CPU firmware to exploit known vulnerabilities in the Enclave implementation. To counter these threats, TeeMAF employs the rollback protection Filesystem from SCONE, which supports the termination of an application's execution upon detecting any rollback of files or individual blocks. Additionally, SCONE prevents firmware rollbacks by verifying the integrity of the platform firmware.

### V. EVALUATION

Two prototypes have been implemented based on the Ethereum smart contract platform for comparison. The first prototype, DROP itself, employs Intel Software Guard Extensions (SGX) technology to enable secure remote attestation using TEE. The second prototype serves as a control group and operates without this attestation mechanism. The performance evaluation focuses on three key parameters that may impact blockchain performance:

1) Number of Blockchain Nodes: The scalability of a blockchain network is intrinsically linked to the number of nodes participating in the network. A higher number of nodes can enhance the security and decentralization of the blockchain but may also introduce latency and synchronization challenges.

2) **Block Construction Time:** This metric refers to the time required to create a new block in the blockchain. The Proof of Authority (PoA) consensus protocol is adopted in the implementation, allowing the block time to be configurable. By adjusting the block time, the system's latency and throughput can be measured.

3) **Transaction Send Rate:** The transaction send rate represents the frequency at which transactions are submitted to the blockchain system. It is a crucial indicator of the network's ability to handle a large volume of transactions and its suitability for high-throughput applications. By varying the transaction send rate, the test results will show whether the confidential computing capabilities and remote attestation mechanisms provided by Intel SGX have an impact on the transaction send rate of the blockchain system.

To quantify the impact of TEE technology on blockchain performance, this paper has meticulously measured latency (seconds) and throughput (transactions per second, TPS), comparing these metrics across scenarios with and without Intel SGX's Remote Attestation (RA). Our experimental setup involved interactions with off-chain functions, both with and without the remote attestation enabled, utilizing the OffChainWorkerRegister() method in the $SC_{SP}$. This paper prepared and processed 1000 transactions to simulate real-world interactions between off-chain and on-chain functions.

### A. Experimental Setup

The experimental apparatus for this study was configured as follows: The primary computing platform for this study was equipped with an Intel® Xeon® E-2286M CPU, which features 16 cores and a base frequency of 2.40 GHz, complemented by 64 GB of RAM, to adequately support the demands of blockchain operations. The study utilized Ubuntu 20.04 LTS as the operating system, providing a stable foundation. The software suite was comprised of Docker Engine (version 25.0.3) for container management, and the Ethereum client Geth (alltools-v1.12.0), developed in Go, which facilitated blockchain interactions. For trusted execution, the study employed Intel® SGX technology, incorporating the Intel® SGX SDK for development, Intel® SGX PSW (version 2.20.100.2) for runtime support, and Intel® SGX DCAP (version 1.20.100.2) for secure attestation, ensuring reliable and secure execution environments. Additionally, the SCONE runtime environment was used, with both the SCONE LAS and CAS being version 5.8.0. A local Ethereum private network has been setup utilizing the PoA consensus protocol, Clique, to investigate the impact of various factors on network performance. The study is particularly focused on analyzing how variations in the number of blockchain nodes and block construction times affect the interaction between off-chain and on-chain functionalities. A constant transaction sending rate is maintained at three different levels (50 tps, 150 tps, and 250 tps), allowing us to measure the resulting latencies and throughput under these conditions.

### B. Blockchain Nodes Impact On Blockchain:

Fig 4. illustrates the impact of the number of blockchain nodes on transaction latency under different sending rates. The study consistently reveals a linear increase in latency as the number of nodes grows, a trend that holds true across all transaction sending rates, regardless of the implementation of a remote attestation mechanism. When the blockchain has only one node and the transaction send rate is 50 tps, the latency gap between the platform with and without the RA mechanism is 16.59%, likely due to the overhead introduced by the attestation processes. Theoretically, the platform with the remote attestation mechanism should experience higher latency due to the additional proof workload introduced, causing greater delays in interactions between off-chain and on-chain functions. As the transaction sending rate escalates to 250 tps, the performance gap narrows to 4.56%, possibly because the relative overhead of the attestation becomes less significant compared to the increased load of transaction processing. As the number of nodes continues to rise, the performance differential further diminishes, achieving a minimal gap of only 0.084%, indicating that at higher scales, the impact of remote attestation on latency becomes negligible.

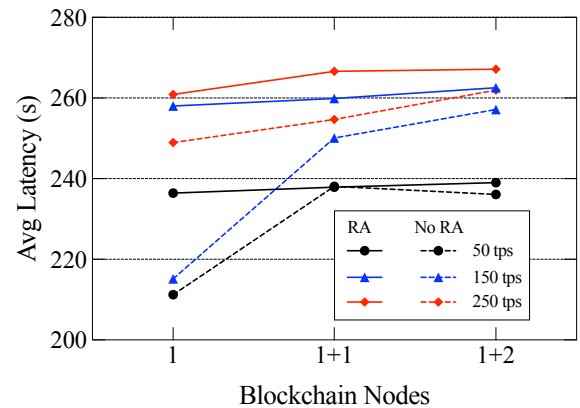

Fig. 4. Impact of Blockchain Nodes on Average Transaction Latency

Fig. 5. illustrates the impact of the number of blockchain nodes on transaction throughput at different sending rates. The study indicates that, with a single node, the maximum difference in transaction throughput between the platform with the RA mechanism and the one without it is 7.14%. At a sending rate of 250 transactions per second, the difference in throughput between the two platforms becomes negligible, suggesting that the transaction processing capacity at higher rates effectively mitigates the overhead introduced by the remote attestation mechanism. As the network expands to three nodes, the maximum difference in transaction throughput declines to just 3.70%, indicating that increased network size helps in distributing the attestation overhead more effectively, thus minimizing its impact on overall transaction processing.

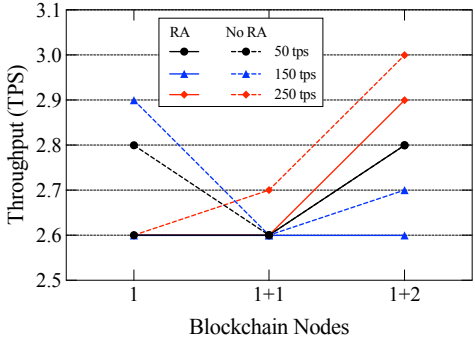

Fig. 5. Impact of Blockchain Nodes on Throughput

### C. Blockchain Construction Time Impact On Blockchain:

Fig. 6. illustrates the relationship between average latency and block time under different transaction sending rates for platform configurations with RA and without RA. It can be observed from the figure that, across all transaction sending rates, the latency exhibits a linear increasing trend as the block time increases. Initially, at a block time of 5 seconds, the latency difference between the platforms with and without remote attestation reaches a maximum of 16.59%. This substantial gap likely results from the additional processing required for RA at lower block times, where its relative impact on transaction handling is more pronounced. As the transaction sending rate increases, the latency for both RA and non-RA platforms rises due to the higher volume of transactions to process. However, the gap between them narrows, suggesting that the overhead from RA becomes less significant compared to the overall latency caused by increased transaction loads. For instance, at 50 tps, the latency of the RA system is consistently slightly lower than that of the non-RA system. This trend persists at higher transaction rates (150 tps and 250 tps), where the performance gap becomes negligible. At a block time of 15 seconds, the average latencies of the RA and non-RA platforms converge, showing only a marginal difference of 3.30%. This convergence indicates that as block times increase, the relative impact of RA on latency diminishes, aligning more closely with the performance of the non-RA platform.

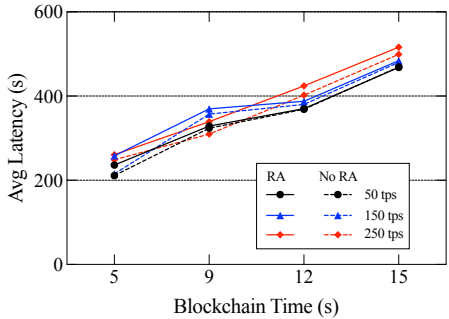

Fig. 6. Impact of Blockchain Time on Average Transaction Latency

Fig. 7. presents the experimental results of blockchain system throughput versus block time (s) for platforms with and without RA at different transaction sending rates (50 tps, 150 tps, and 250 tps). The figure illustrates a clear trend where, as block time increases, throughput consistently decreases across all sending rates and both platform configurations. This is typically because longer block times delay the confirmation of transactions, reducing the number of transactions that can be processed per unit time. At an initial block time of 5 seconds, throughput peaks, showcasing minimal difference between platforms with and without RA, with a variance of just 7.14%. This suggests that at faster block times, the overhead introduced by RA is less impactful on the overall system throughput. At a block time of 9 seconds, the throughput difference slightly narrows to 6.25%, indicating that as block times extend, the relative efficiency of both configurations begins to align more closely, despite the added security processes of RA. As block time further increases to 12 and 15 seconds, throughput significantly diminishes across all scenarios, and intriguingly, the performance curves of platforms with and without RA align almost parallelly. This alignment suggests that at longer block times, the transaction processing capacity becomes the limiting factor, overshadowing any latency differences introduced by RA.

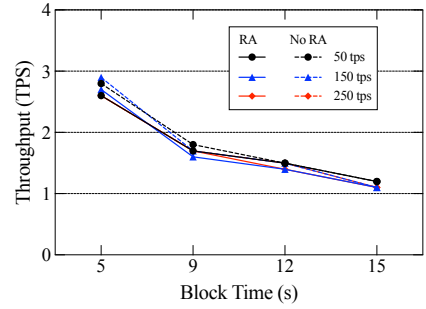

Fig. 7. Impact of Blockchain Time on Throughput

### D. Discussion

The results obtained by varying the number of blockchain nodes and block time both indicate that the platform with the remote attestation mechanism exhibits a performance overhead compared to the platform without this mechanism. The experimental analysis investigates this impact by testing different transaction sending rates and observing the average latency and throughput over 1000 transaction rounds.

The analysis reveals that while the RA mechanism initially introduces additional latency, this impact lessens with increased transaction loads and a higher number of nodes. This diminishing effect underscores the RA mechanism's scalability and its capability to maintain efficiency under high-demand conditions, proving its suitability for complex, large-scale blockchain networks. As the block time increases, the latency gap between the platform with the RA mechanism and the one without gradually decreases. The slower block time allows the entire system more time to process and validate each transaction, which is applicable to complex smart contracts and applications. Therefore, compared to the platform without the RA mechanism, the platform with the RA mechanism can achieve virtually indistinguishable latency in a stable system. Concerning transaction throughput, the performance differential between the platforms narrows to an acceptable range as the transaction load and node count increase. This acceptable range indicates that the RA mechanism's overhead becomes less significant in larger, busier systems, aligning more closely with the throughput levels of the non-RA platform. Moreover, as block time extends, the additional overhead from the RA mechanism becomes increasingly negligible in affecting throughput, which reinforces that block time is the predominant factor dictating overall performance. Longer block times allow the system ample opportunity to process and validate transactions efficiently, even in the presence of complex security mechanisms like RA.

## VI. RELATED WORK

This section presents work related to the framework that focuses on blockchain off-chain trusted computing.

### A. Secure Multi-Party Computation

Due to the demand for off-chain trusted computing, researchers are attempting to deploy secure Multi-Party Computation (MPC)s on the blockchain to address privacy concerns. Zyskind et al. [30] introduced the Enigma platform, which leverages secure MPC to decentralize data processing across network nodes, enhancing privacy by fragmenting datasets so no single node has complete data access. This method significantly mitigates the risk of data exposure. However, this fragmentation introduces complexities in coordinating decentralized computations and node roles, potentially complicating the scalability and operational efficiency of the platform.

David Cerezo Sánchez [31] introduced the Raziel system, which integrates proof-carrying code with MPC in smart contracts to ensure confidential and verifiable execution. This effectively addresses both Gyges and DAO attacks. Raziel enhances the security of smart contracts by utilizing non-interactive zero-knowledge proofs to verify computations without exposing the underlying data or logic. The system proposes a model in which miners are compensated for generating pre-processing data for secure MPC. Raziel introduces mechanisms to reduce computational costs through outsourcing to miners and pre-processing. However, scalability in larger or public blockchain environments remains a significant challenge.

Zhou et al. [32] investigate the integration of MPC into the Hyperledger Fabric blockchain. They utilize homomorphic encryption and secret sharing to secure sensitive transaction data on permissioned blockchains. This protocol enables publicly verifiable computations, which enhances transparency and trust in processed transactions. Despite these advances, the approach suffers from significant communication overhead, which affects the scalability and efficiency of the system, potentially limiting its practical application in larger or more complex network environments. Benhamouda et al. [33] addressed providing private data options on Hyperledger Fabric. They integrated secure MPC protocols directly into the blockchain as part of the smart contract, utilizing the blockchain's built-in facilities for identity verification and communication.

### B. Zero-Knowledge Proof

Zero-knowledge proofs (ZKPs) have emerged as a promising technique to address the challenges of privacy and scalability in blockchain systems. ZKPs allow a prover to convince a verifier about the validity of a statement without revealing any additional information beyond the statement's validity [34]. Bulletproofs [35] is a prominent ZKP scheme that provides short and computationally efficient proofs, making it suitable for blockchain applications.

To enable privacy-preserving off-chain computations, ZoKrates [36] introduces a toolbox that allows developers to specify, integrate, and deploy off-chain computations using zk-SNARKs. ZoKrates provides a domain-specific language, a compiler, and generators for proofs and verification smart contracts, abstracting away the complexities of ZKPs. This enables developers to write provable off-chain programs without deep knowledge of the underlying proof system. zkrpChain [37] proposes a range-based privacy-preserving data auditing solution for consortium blockchains. It supports the generation and verification of zero-knowledge range proofs for both standard and arbitrary ranges, as well as the aggregation of multiple proofs and batch verification. zk-AuthFeed [38] addresses the limitation of Town Crier, which relies on a trusted execution environment for authenticated data feed to smart contracts. zk-AuthFeed combines zk-SNARKs with digital signatures to achieve both data authenticity and privacy for off-chain data fed to smart contracts. zk-Oracle [39] proposes batching techniques for proof generation to improve the efficiency and scalability of zk-proof generation. The approach leverages horizontal and vertical batching patterns, significantly reducing the time required for zk-proof generation while optimizing the proof size to minimize on-chain verification costs.

In the domain of blockchain-based e-voting systems, Emami et al. [40] propose a scalable decentralized privacy-preserving e-voting system based on zero-knowledge off-chain computations. The system employs zk-SNARKs and introduces the concept of "ballot boxes" to offset voters' transaction costs. The scheme promotes transparency, privacy, universal verifiability, and weak receipt-freeness, making it suitable for large-scale elections.

### C. Trusted Execution Environments

Trusted Execution Environments (TEEs), such as Intel SGX, have been explored in conjunction with blockchains to enable secure off-chain computation and storage while maintaining the decentralized trust properties of blockchains.

Brandenburger et al. [41] discuss the challenges and conceptual differences of on-chain and off-chain execution when combining smart contracts with trusted computing. They highlight the need to carefully design systems that leverage TEEs and blockchains to avoid canceling out their respective advantages. Similarly, Dang et al. [42] propose using TEEs to scale blockchains by offloading computation and storage to SGX-enabled nodes while ensuring the integrity and confidentiality of off-chain data.

Several works have focused on utilizing TEEs for specific off-chain functionalities. Lind et al. [43] introduce Teechain, an off-chain payment network that uses TEEs to secure funds in payment channels without requiring synchronous access to the blockchain. Cheng et al. [44] propose a lightweight mobile client privacy protection method based on TEE and blockchain (LMCPTEE) to enhance the privacy of light clients. Matetic et al. [45] present BITE, an approach to protect the privacy of light clients in Bitcoin using enclaves in full nodes.

In the context of IoT, off-chain trusted computing has been explored to ensure the integrity and confidentiality of IoT data. Xie et al. [46] propose TEDA, a trusted execution environment-and-blockchain-based data protection architecture for IoT. TEDA uses SGX to process local data and manage device identities, while leveraging blockchain for secure access control and verification. Similarly, Xie et al. [47] introduce TEBDS, a TEE-and-blockchain-supported IoT data sharing system that combines on-chain and off-chain methods to meet the security requirements of IoT data sharing.

Zhang [48] presents Truxen, a trusted computing enhanced blockchain that uses a Proof of Integrity protocol as the consensus mechanism. Truxen leverages trusted computing to enhance the areas of mining blocks, executing transactions and smart contracts, and protecting sensitive data. The proposed Single Execution Model allows transactions and smart contracts to be verified and executed on a single node, enabling remote calls to off-chain applications and supporting in-deterministic tasks. Cheng et al. [49] introduce Ekiden, a platform for confidentiality-preserving, trustworthy, and performant smart contract execution. Ekiden offloads smart contract execution from the blockchain to TEEs, enabling thousands of transactions per second while preserving data confidentiality and integrity. Brandenburger et al. [50] present an architecture and prototype for smart contract execution within Intel SGX for Hyperledger Fabric. Their system resolves difficulties posed by Fabric's execute-order-validate architecture, prevents rollback attacks on TEE-based execution, and minimizes the trusted computing base by encapsulating each application within its own enclave.

## VII. CONCLUSION

This paper addresses the security and integrity issues of deploying applications through blockchain in edge or cloud environments by proposing a scheme called TeeMAF. The framework achieves mutual attestation between on-chain and off-chain functions in DApps. Additionally, it proposes a decentralized resource orchestration platform based on TeeMAF for users to deploy the off-chain functions of DApps. The core concept in the approach is to establish trust through the remote attestation mechanism of Intel SGX technology in TEE and to enable on-chain functions to prove the corresponding off-chain functions through each remote attestation. The experimental section has conducted validation in a local deployment of the Ethereum platform and performed blockchain system performance experiments using Hyperledger Caliper. The results demonstrate that TeeMAF can effectively maintain security and reliability with minimal performance overhead and can scale to larger blockchain networks. Future work will investigate how to further optimize the Intel SGX remote attestation process in TeeMAF, reduce attestation latency, and improve attestation efficiency.

ACKNOWLEDGMENT

This work was supported in by Science Foundation Ireland (SFI) under Grant Number SFI 22/CC/11147, co-funded by the European Regional Development Fund.